\documentclass[a4paper,11pt]{article}
\pdfoutput=1 
\usepackage{jheppub} 
\usepackage[T1]{fontenc} 
\usepackage{subcaption}

\usepackage[normalem]{ulem}
\usepackage[dvipsnames]{xcolor}

\title{\boldmath Model selection and signal extraction using Gaussian Process regression}

\author[a]{Abhijith Gandrakota,}
\author[b]{Amitabh Lath,}
\author[b]{Alexandre V. Morozov,}
\author[c]{Sindhu Murthy}
\affiliation[a]{Fermi National Accelerator Laboratory, Batavia, USA}
\affiliation[b]{Department of Physics \& Astronomy, Rutgers, The State University of New Jersey,\\ Piscataway, USA}
\affiliation[c]{Department of Physics, Carnegie Mellon University, Pittsburgh, USA}
\emailAdd{abhijith@fnal.gov}
\emailAdd{lath@physics.rutgers.edu}
\emailAdd{morozov@physics.rutgers.edu}
\emailAdd{sindhum@andrew.cmu.edu}

\abstract{We present a novel computational approach for extracting weak signals, whose exact location and width may be unknown, from complex background distributions with an arbitrary
functional form. We focus on datasets that can be naturally presented as binned integer counts, demonstrating our approach on
the CERN open dataset from the ATLAS collaboration at the Large Hadron Collider, which contains the Higgs boson signature.
Our approach is based on Gaussian Process (GP) regression - a powerful and flexible machine learning technique that allowed us to model the background without
specifying its functional form explicitly, and to separate the background and signal contributions in a robust and reproducible manner. Unlike functional fits, our GP-regression-based approach
does not need to be constantly updated as more data becomes available. We discuss how to select the GP kernel type, considering trade-offs between kernel complexity and its ability to capture
the features of the background distribution. We show that our GP framework can be used to detect the Higgs boson resonance in the data with more statistical significance than a polynomial fit specifically tailored to the dataset.
Finally, we use Markov Chain Monte Carlo (MCMC) sampling to confirm the statistical significance of the extracted Higgs signature.
}

\begin{document} 
\maketitle
\flushbottom
\section{Introduction}
Analyzing data from physical experiments or observations often involves fitting computational models in order to extract a signal in the presence of both background effects and random noise. For example, such a setting appears naturally in the analysis of X-ray diffraction patterns from crystalline samples that contain contributions both from distinctive Bragg peaks and diffuse background scattering~\cite{Sivia1994,David2001}, inference of transiting exoplanet parameters in astronomy~\cite{exoplanet_astro,gp_astro_general},
and the discovery of the Higgs boson and search for the new physics at the Large Hadron Collider (LHC) at CERN~\cite{Aad_2012}. The data from LHC and similar experiments usually comes in the form of binned integer counts~\cite{Zyla:2020zbs}. Traditionally, modeling such data is performed under the assumption of the Poisson distribution, by employing a parametric fit~\cite{Zyla:2020zbs}. The fitted models are subsequently used to estimate the background contributions and extract the signal of interest. However, the choice of parametric functions is often ad-hoc and the degree of model complexity requires a delicate balancing act between overfitting and underfitting the data~\cite{bishop2006pattern,Mehta2019,Rocks2020}.

For data analyses of the type performed on the LHC bin counts, the optimal complexity of the model is usually evaluated by performing Wilk's tests~\cite{Wilks1938}. Most analyses that employ this technique test it on a fraction of the full dataset, usually 10\% or less. This process is called ``blinding'' and is meant to reduce biases in the analysis. However, the model selection process must be repeated periodically as more data becomes available, and often the functional form employed in the model has to be updated as well~\cite{Aad_2016}. Using non-parametric methods such as Gaussian Process regression is an effective way of alleviating these concerns~\cite{pmlr-v5-titsias09a,bishop2006pattern,gpml,Kersting2007,ohagan}.

Gaussian Process (GP) regression is a well-established machine learning technique~\cite{gpml,bishop2006pattern} commonly used in various fields such as astrophysics, gravitational wave detection, and high energy physics~\cite{Iyer_2019,Moore_2016,golchi2015,frate2017modeling}. In particular, in high energy physics GP regression was used to model the smooth continuum background from quantum chromodynamics (QCD) in searches for dijet resonances in LHC data~\cite{frate2017modeling}. The authors argued that using GP for background estimation was more robust with respect to increasing luminosity compared to parametric fitting methods. 
GP regression's advantages over more conventional methods, which employ a linear expansion over a fixed set of basis functions such as polynomials or Gaussians, are due to its non-parametric flexibility and a principled Bayesian framework. Instead of explicit basis functions, GP regression is defined in terms of kernel functions which specify the degree of correlations between two points in the dataset. The GP approach allows us to perform inference using a much broader class of functions, including those which would otherwise require an infinite basis set~\cite{bishop2006pattern}.
GP regression is robust with respect to the size of the dataset~\cite{frate2017modeling}.

Nevertheless, the flexibility of GP regression can be a double-edged sword. In GP regression, kernel functions typically depend on several hyperparameters that are varied to fit the data, typically through non-Bayesian techniques such as maximizing the marginal likelihood of the observed data~\cite{gpml,bishop2006pattern}. The hyperparameters describing the kernel function control the flexibility of the resulting model, while the type of the kernel function determines the success in capturing certain features in the data, such as periodic oscillations and long-term trends~\cite{gpml}. Thus, the universality and the power of the GP approach may come at the cost of overfitting with respect to both kernel type and kernel hyperparameter choices. Therefore, a method is required that can constrain the flexibility of the GP regression in a controlled manner.

Previous work in this area has focused on ``kernel learning'' to address the issues of flexibility and robustness, with several techniques proposed that aim at constructing composite kernels for Support Vector Machines~\cite{Diosan2007EvolvingKF}, Relevance Vector Machines~\cite{Bing2010}, and GP regression~\cite{duvenaud2013structure,Wilson2013} using a library of base kernels.
Semiparametric regression attempts to combine interpretability of parametric models with flexibility of non-parametric models by combining the two approaches in a single framework~\cite{Ruppert2003}. However, none of the above approaches focus specifically on integer count data or on the processes that are naturally viewed as localized signals superimposed on the smooth background.
A previous application of GP regression to LHC data~\cite{frate2017modeling} employed both standard and custom-built kernels motivated by physical considerations.
In contrast, in this work we have developed both a model selection procedure suitable for GP regression and an approach for estimating the statistical significance of the extracted signal.

New signals in physical observations of particle resonances in LHC data often appear as localized features (``bumps'') superimposed on a smooth background. Accurate modeling of the background spectrum is therefore essential to both extracting the signal and assessing its statistical significance.
In this paper, we present a rigorous approach to model selection in GP regression applied to binned integer data, which we expect to be a superposition of a localized signal and a smooth background of unknown functional form. We exploit the flexibility of the GP regression by determining the kernel hyperparameters through the fit to background-only data, with the signal window masked out.
These parameters are subsequently used to extrapolate the background contribution across the signal window, enabling us to separate the background from the signal contribution.
We describe procedures for kernel type selection based on both Bayesian and Akaike information criteria.
We also propose a method for estimating the statistical significance of the signal by performing a hypothesis test with data \textit{devoid of signal} as the null hypothesis and data \textit{containing signal} as the alternate hypothesis. While similar in spirit to standard hypothesis-testing approaches,
our significance test takes into account both the uncertainties inherent in the Bayesian nature of GP regression and the sampling noise related to
generating integer bin counts from GP-predicted, real-valued Poisson rates.

In this work, we illustrate our procedure by detecting for the Higgs boson resonance in the open data collected by the ATLAS experiment at the LHC ~\cite{ATL-OREACH-PUB-2020-001}.
We show that using GP regression leads to extracting the Higgs boson signature at a higher level of statistical significance compared to parametric fits.
Our computational pipeline can be applied for background estimation and signal detection in any dataset where a localized signal is obscured by background processes.

\section{Model selection and signal extraction procedure}

\subsection{Gaussian process regression}

In regression, the observed data $y = (y_1 \dots y_N)$ (in our case, integer counts in $N$ bins)
is modeled by $z = (z(x_1) \dots z(x_N))$:
\begin{equation} \label{regression}
    y_k = z(x_k) + \epsilon_k,
\end{equation}
where $k = 1 \dots N$ ($N$ is the total number of datapoints), $X = (x_1 \dots x_N)$ is a vector of input variables (in our case, centers of the bins with integer counts), and $\epsilon_k$ is a random noise variable independently sampled from a Gaussian distribution ${\cal N} (\epsilon | 0,\sigma^2_i)$ for each data point, where $\sigma^2_i$ is the noise variance in bin $i$.

In the GP framework, the model $z(x)$ is not represented as an explicit linear expansion over a set of pre-determined basis functions. Instead, we directly consider the marginal likelihood $p(y|X)$ integrated over all possible models~\cite{Goldberg1998,bishop2006pattern}:
\begin{equation} \label{GP:marginal}
    p(y|X) = \int dy~p(y|z) p(z|X) = {\cal N} (y|m,C),
\end{equation}

where $m = (m(x_1) \dots m(x_N))$ is a vector of values of the mean function $m(x)$ for all datapoints, and the covariance matrix $C = K + \Sigma$, where $K$ is the Gram matrix and $\Sigma$ is a diagonal $N \times N$ matrix with $\Sigma_{ii} = \sigma^2_i$. 
The elements of the Gram matrix are values of the kernel function $k(x,x')$ evaluated for all pairs of input variables: $K_{ij} = k(x_i,x_j)$. Note that, consistent with Eq.~\eqref{regression}, $p(y|z) = {\cal N} (y|z,\Sigma)$, while
$p(z|X) = {\cal N} (z|m,K)$ from the definition of the Gaussian Process.

Thus, GP regression is defined by the mean function $m(x)$ and the kernel function $k(x,x')$~\cite{gpml,bishop2006pattern}, which determines the degree of correlation between any two datapoints. In general, the kernel function depends on a set of $n$ hyperparameters $\theta = (\theta_1, \theta_2, \dots , \theta_n)$, whose number and meaning depend on the kernel type. The hyperparameters of a given kernel are usually optimized by maximizing the marginal likelihood in Eq.~\eqref{GP:marginal}, a non-Bayesian procedure~\cite{gpml,bishop2006pattern}. Ordinarily,
kernel hyperparameters include $\sigma^2_i$, which represent the amount of experimental noise in each bin. However, since this would introduce too many hyperparameters and make
their optimization difficult or impossible, we estimate $\sigma^2_i$ directly from the data using Garwood intervals which allow us to extract two-sided confidence intervals from
the number of events in each bin under the assumption that the events are Poisson-distributed~\cite{Garwood1936,Demortier2013}. Thus, $\sigma^2_i$ are estimated independently and are
not treated as hyperparameters in our approach.

The strength of the GP approach stems from the fact that the joint marginal probability of observing a set of datapoints is Gaussian (Eq.~\eqref{GP:marginal}). Moreover, the predictive probability $p(\tilde{y}_i|y)$, the conditional probability distribution of observing a real-valued ``count'' $\tilde{y}_i$ in bin $i$ given a dataset with $N$ previous observations, is also Gaussian, with the mean $f(x_i)$ and the variance $V(x_i)$ given by:
\begin{eqnarray} \label{mean:var:general}
    f(x_i) &=& m(x_i) + \tilde{k}^T C^{-1} \tilde{y}_i, \\
    V(x_i) &=& \alpha - \tilde{k}^T C^{-1} \tilde{k}, \nonumber
\end{eqnarray}
where $\tilde{k} = (k(x_1,x_i), \dots ,k(x_N,x_i))$ and $\alpha = k(x_i,x_i) + \sigma^2_i$. \\

In this work, we consider two types of GP regression: with $m(x_i) = 0,~\forall i$ for modeling the background-only distribution, and with the Gaussian mean function for modeling signal+background datasets, where the signal component is represented by:

\begin{equation} \label{GP:meanf}
    m(x_i) = \frac{A}{\sqrt{2 \pi} \sigma} \exp \left(- \frac{ \left(x_i - \mu \right)^2 }{2\sigma^2} \right).
\end{equation}
Here, $A$ defines the signal strength, while $\mu$ and $\sigma$ represent signal mean and width, respectively.
The rounded value of $A$ can be interpreted as the total number of signal events. Note that when the Gaussian mean function is introduced,
the set of model hyperparameters $\theta$ needs to be augmented with $\{ A, \mu, \sigma \}$.





\subsection{Model selection}

A key issue in GP regression is the choice of a kernel and, given a kernel, derivation of the optimal set of hyperparameters $\hat{\theta}$ for it. Typically, the optimal set of hyperparameters
is obtained by maximizing the marginal log-likelihood $\log p(y|X,\theta,K_i)$~\cite{gpml,bishop2006pattern}, where $p(y|X,\theta,K_i)$ is given by Eq.~\eqref{GP:marginal} and its dependence on the set of hyperparameters $\theta$ and the kernel type $K_i$ are made explicit for clarity.
Since this step is non-Bayesian, a question of kernel selection arises which would take into account both kernel complexity (i.e., the amount of signal smoothing provided by a given kernel) and the number of kernel hyperparameters. A standard way for carrying out model comparison is based on the 
Bayesian Information Criterion (BIC) for the marginal log-likelihood~\cite{lhr,bishop2006pattern}:

\begin{equation} \label{eq:bic}
    - \log p(y|X,K_i) \simeq - \log p(y|X,\hat{\theta},K_i) + \frac{n}{2} \log N \equiv \text{BIC}^\text{naive},
\end{equation}
where $p(y|X,K_i)$ is the model evidence (likelihood marginalized over hyperparameters), $n$ is the number of model parameters, and $N$ is the number of datapoints. Note that the second term on the right-hand side penalizes model complexity, such that lower BIC scores are more preferable. The derivation of BIC relies on a number of approximations whose validity depends on the details of the system under consideration. Specifically, the derivation employs the Laplace approximation to estimate the integral over the hyperparameters and assumes that $N$ is so large (or the Gaussian prior distribution over the hyperparameters is so broad) that the effects of the hyperparameter priors are negligible, resulting in:

\begin{equation} \label{eq:bic:full}
    - \log p(y|X,K_i) \simeq -\log p(y|X,\hat{\theta},K_i) + \frac{1}{2} \log |H| \equiv \text{BIC},
\end{equation}
where $H = - \vec{\nabla}_\theta \vec{\nabla}_\theta \log p(y|X,\theta,K_i) \vert_{\hat{\theta}}$ is the Hessian in the model hyperparameter space evaluated at the hyperparameter values that maximize
the marginal log-likelihood. If $N$ is large and the Hessian has full rank, the second term on the right-hand side can be roughly approximated as $\frac{1}{2} \log |H| \simeq \frac{n}{2} \log N$, yielding Eq.~\eqref{eq:bic}.

An elegant alternative approach to model selection is based on the Akaike Information Criterion (AIC), which accounts for the fact that the log-likelihood computed on a training dataset
provides an estimate of the prediction error that is too optimistic, because the same data is being used to fit the model and assess its error~\cite{Akaike1974,hastie2001elements}.
To account for this optimism, a correction term is added which is based on the sum of covariances between the observed datapoint and the newly generated datapoint for each
input variable $x_i$. It can be shown that the sum of covariances is proportional to the number of degrees of freedom in the $N \to \infty$ limit, resulting in the following expression for AIC:
\begin{equation} \label{eq:aic}
     \text{AIC} \equiv - 2 \log p(y|X,\hat{\theta},K_i) + 2 d,
\end{equation}
where $d$ is the number of degrees of freedom in the model. Thus AIC provides an estimate of the log-likelihood that would have resulted if another dataset was independently generated at the
same values of input variables (an in-sample estimate). In the case of GP regression, $d$ needs to be replaced by $d_{\mathit{eff}}$ in Eq.~\eqref{eq:aic}, where $d_{\mathit{eff}}$ is the effective number of degrees of freedom for the GP regression with a given kernel type, which captures the amount of smoothing induced by the GP fit~\cite{gpml,gam}:

\begin{equation} \label{dof}
    d_{\mathit{eff}} (\hat{\theta}) = \mathrm{tr} [K(\hat{\theta})(K(\hat{\theta}) + \Sigma)^{-1}],
\end{equation}
where the dependence of the Gram matrix on the optimal kernel hyperparameters $\hat{\theta}$ is made explicit for clarity. Note that similar to BIC, lower AIC values are preferable;
however, unlike BIC, AIC is a non-Bayesian measure and thus provides an alternative approach to model selection.

Choosing the appropriate kernel type is crucial to the success of GP regression, since different kernels emphasize different correlation structures in the data. In practice, kernels are often constructed manually using simple comparison metrics such as marginal likelihood or BIC$^\text{naive}$. In some cases, composite kernels are constructed automatically using kernel engineering techniques (see e.g. Ref.~\cite{duvenaud2013structure}).
Here, we propose a kernel selection technique which is based on the consensus between AIC and BIC measures of model complexity.
This framework allows us to compare models with different kernels and choose a specific kernel type on the basis of both a Bayesian approach to model selection, which emphasizes the complexity
of the kernel in terms of the number of kernel parameters, and a non-Bayesian approach, which is based on the amount of smoothing introduced by the GP fit.

\subsection{Poisson likelihood}

Since our data consists of integer counts in $N$ bins, we have also employed a Poisson-type model to generate integer predictions in each bin. Specifically, we assume that the mean
of the predictive probability $f(x_i)$ (Eq.~\eqref{mean:var:general}) provides the rate for the Poisson process in each bin~\cite{frate2017modeling}:
\begin{equation} \label{pdg}
   \log{L_{\cal P}} = \sum_{i=1}^N \left[ y_i - f(x_i) - y_i \log{\left( \frac {y_i} {f(x_i)} \right)} \right].
\end{equation}
Note that $f(x_i)$ implicitly depends on the kernel type and the optimized hyperparameter values $\hat{\theta}$.
Eq.~\eqref{pdg} can be used both to generate integer counts and compute the log-likelihood of the observed counts.
Poisson log-likelihood can also be used instead of the GP marginal log-likelihood to compute BIC (Eq.~\eqref{eq:bic:full}), $\text{BIC}^\text{naive}$ (Eq.~\eqref{eq:bic}),
and AIC (Eq.~\eqref{eq:aic}).

\subsection{Gaussian process kernels}

We used the GP package from scikit-learn (\texttt{https://scikit-learn.org/stable/}), augmenting the implementation to include custom kernels and kernel extraction features. In this paper, we have explored three kernels to model the continuum background distribution: the Radial Basis Function kernel (RBF), the Mat{\'e}rn kernel with $\nu = 5/2$ (Matern),
and the second-order polynomial kernel (Poly2). The three kernel functions are defined below:



\begin{equation} \label{RBF}
    k_{\text{RBF}} (x,x') = \sigma_0 \exp \left[ - \frac{ \left(x-x'\right)^2 }{2 l^2}\right],
\end{equation}
where $\sigma_0$ is the amplitude and $l$ is the length scale of the covariance function;

\begin{equation} \label{matern}
    k_{\text{Matern}} (x,x') = \sigma_0 \Bigg[1 + \frac{\sqrt{5}}{l} d(x , x' ) +\frac{5}{3l} d(x , x' )^2 \Bigg] \exp \Bigg[-\frac{\sqrt{5}}{l} d(x , x' ) \Bigg],
\end{equation}
where $\sigma_0$ is the covariance amplitude as in $k_{\text{RBF}}$, $l$ is a positive parameter characterizing the covariance, and $d(x,x')$ is the Euclidean distance between datapoints $x$ and $x'$;

\begin{equation} \label{poly}
    k_{\text{Poly2}} (x,x') = (\sigma_0^2 + x \cdot x')^2,
\end{equation}
where $\sigma_0$ sets the magnitude of the zeroth-order term in the polynomial expansion. Thus the RBF, Matern and Poly2 kernels depend on 2, 2 and 1 hyperparameter, respectively.



\subsection{Functional fit}

For comparison, we also employ a fourth-order parametric polynomial fit with explicit basis functions, which is typically used to model the background distribution~\cite{ATL-OREACH-PUB-2020-001}:
\begin{equation} \label{eq:func4}
f(x_i) = \sum_{p=0}^{4} w_p x_i^p + m(x_i),
\end{equation}
where $w_p$ are the fitting coefficients and $m(x_i)$ is either set to $0$ for background-only fits with the signal window masked out, or given by Eq.~\eqref{GP:meanf} for signal+background fits on the entire dataset.
The fits were carried out using ROOT data analysis software~\cite{Brun1997}, by maximizing the Poisson log-likelihood in Eq.~\eqref{pdg}.
For the background-only fit, the values of the fitting coefficients are $w_0 = 1.84 \times 10^5 \pm 1.60 \times 10^2$, $w_1 = -4.49 \times 10^3 \pm 1.80 \times 10^0$,
$w_2 = 4.22 \times 10^1 \pm 1.00 \times 10^{-2}$,  $w_3 = -1.79\times10^{-1} \pm 9.00\times10^{-4}$, $w_4 = 2.84\times10^{-4} \pm 4.0\times10^{-7}$.
For the background+signal fit, the fitting coefficients are $w_0 = 1.64 \times 10^5 \pm 6.16 \times 10^4$, $w_1 = -3.87 \times 10^3 \pm 1.86 \times 10^3$,
$w_2 = 3.52 \times 10^1 \pm 2.10 \times 10^1$, $w_3 = -1.43 \times 10^{-1}\pm 1.05 \times 10^{-1}$, $w_4 = 2.19 \times 10^{-4} \pm 1.94 \times 10^{-4}$,
while the signal contribution is described using
$\{ A, \mu, \sigma \} = \{ 443 \pm 199, 124.5 \pm 0.8, 2.3 \pm 0.9 \}$. All the parameter uncertainties have been estimated via Hessian analysis available in ROOT. The value of $A$ and its
uncertainty have been rounded to correspond to the integer number of events. The datasets on which the fits have been performed are described in more detail below.

\section{Datasets}

We use the di-photon sample from the open dataset made available by the ATLAS collaboration at LHC~\cite{ATL-OREACH-PUB-2020-001}. We use the selection criteria as documented in Ref.~\cite{ATL-OREACH-PUB-2020-001} to create a di-photon invariant mass distribution, $m_{\gamma \gamma}$, that shows the Higgs decay.
The Higgs decay is a localized bump on top of the smooth background distribution, traditionally modeled by a polynomial~\cite{frate2017modeling}. 
The di-photon distribution consists of integer event counts $y_i$ in $N=30$ bins.
Since in this work we focus on the datasets in which we expect to find a localized signal whose location is approximately known, we first mask out the region containing the signal.
We expect the signal to be localized with a characteristic width that is small compared to the characteristic length scale describing the background shape~\cite{frate2017modeling}. In new resonance searches, we typically scan for the signal at multiple points within the full range of the dataset, with a prior expectation for the signal width.

To search for a signal in a specific window, we use the entire range of data with the signal window masked out to determine the optimal parameters for the background-only GP regression fit. In new resonance searches,
this process could be repeated for multiple masked-out signal windows. Here, we model the signal using a simple Gaussian whose mean $\mu$ and width $\sigma$ are approximately known to be 125 GeV and 2.5 GeV, respectively~\cite{ATL-OREACH-PUB-2020-001,Aad_2012}. Thus, in the background-only fits we mask out a signal window $\pm 2 \sigma$ around the signal mean $\mu$;
all data outside of this window are assumed to belong to the background distribution and are therefore fit using GP regression with $m(x_i) = 0$. 
Specifically, we optimize the parameters $\theta$ of a given kernel by maximizing the marginal log-likelihood (Eq.~\eqref{GP:marginal}),
which yields the optimal set of hyperparameters $\hat{\theta}$. Given $\hat{\theta}$, the predictive distribution is then provided by Eq.~\eqref{mean:var:general} with $m(x_i) = 0$.
The resulting hyperparameter values are $\{ \sigma_0, l \} = \{ 5.39 \times 10^7, 46.0 \}$ for the RBF kernel, $\{ \sigma_0, l \} = \{ 1.12 \times 10^8, 124.0 \}$ for the Matern kernel, and
$\sigma_0 = 29.3$ for the Poly2 kernel.

\section{Model selection for background-only fits}

To determine which kernel type best represents our data, we have carried out model selection using BIC and AIC model comparison measures, summarized in Table~\ref{kers}.
We find that the marginal log-likelihood is much worse for Poly2 than for either RBF or Matern. This disadvantage is too substantial to be offset by the fact that Poly2 uses one less hyperparameter.
As a result, both $\text{BIC}^{\text{naive}}$ (Eq.~\eqref{eq:bic}) and $\text{BIC}$ (Eq.~\eqref{eq:bic:full}) rank the kernels in the same way, giving a slight edge to RBF over Matern.
Note that this rank is the same as with the marginal log-likelihood without BIC corrections.
However, Matern is slightly favored over RBF when considering Poisson log-likelihoods with rates provided by the mean of the GP predictive distribution (Eq.~\eqref{pdg}).
This preference for Matern holds when the Poisson log-likelihoods are augmented with complexity corrections to produce $\text{BIC}^{\text{naive}}$ scores, while Func4 becomes strongly disfavored
due to its larger number of fitting parameters.

To investigate this matter further, we have considered the effect of the AIC penalty, which
effectively accounts for the amount of smoothing effected by each kernel type~\cite{gam,Buja1989}. We observe that RBF is favored over Matern when the AIC penalty is taken into account
(Table~\ref{kers}). Furthermore, Poisson log-likelihoods slightly favor Func4 over RBF or Matern GP fits.
However, this slight advantage disappears when the AIC correction is taken into account, with the best score assigned to GP regression with the RBF kernel (Table~\ref{kers}).
Overall, we conclude that with Poisson log-likelihoods, there is a slight advantage for RBF over Func4 on the basis of AIC
and a distinct advantage for RBF or Matern over Func4 on the basis of $\text{BIC}^{\text{naive}}$.
Considering all the evidence together, it appears that GP regression with the RBF kernel is the best way to model our data, although the preference of RBF over Matern is fairly slight. 

\begin{figure}[htpb]
    \centering
    \includegraphics[width=9.5cm]{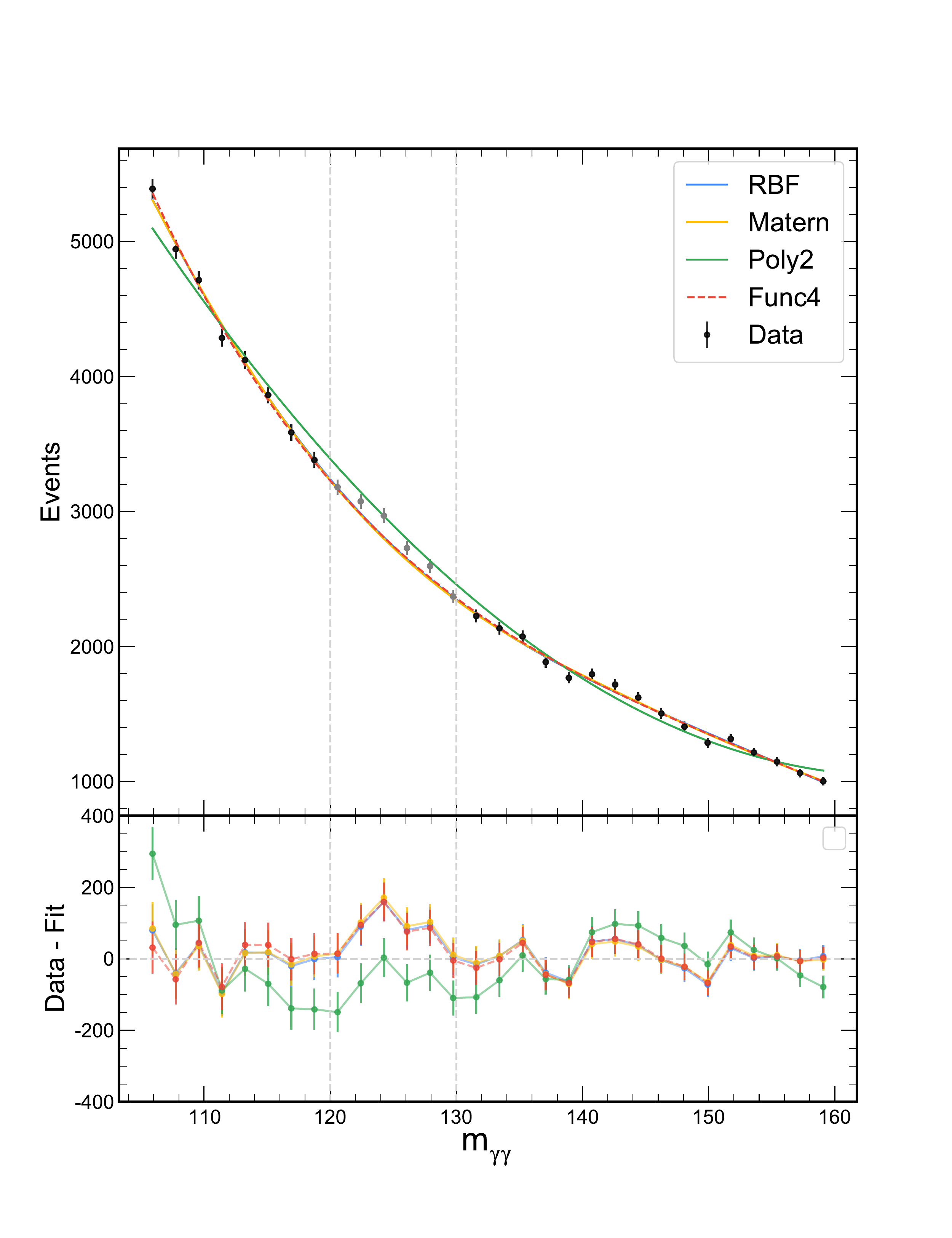}
    \caption{\textbf{Background distribution fits}. Upper panel: Shown are the mean predictive distributions predicted by GP regression with RBF, Matern and Poly2 kernels, as well as the ML fit with the fourth-order polynomial (Func4). All models were fit on the background-only data outside of the $125 \pm 5 \text{ GeV}$ window where the signal is located. Event counts $y_i$ are shown with error bars constructed using Garwood intervals (black markers: background only, grey markers: background+signal). Lower panel: Residuals $y_i - f(x_i)$ for each of the four models in the upper panels, with 
$y_i$ error bars from the upper panel. Residual values are connected by lines of the same color and type as in the upper panel, to guide the eye. In both panels, vertical dashed lines indicate the boundaries of the signal
window. 
    }
    \label{fig:lrt_new}
\end{figure}

In addition to the AIC and BIC-based model selection, we have carried out visual comparisons of the four different models, by plotting the mean predictive distribution of the GP regression
with RBF, Matern and Poly2 kernels ($f(x_i)$ in Eq.~\eqref{mean:var:general} with $m(x_i) = 0$), and the maximum-likelihood (ML) Func4 fit (Eq.~\eqref{eq:func4}) to the event counts outside of
the signal window (Fig.~\ref{fig:lrt_new}). It is clear from the upper panel of Fig.~\ref{fig:lrt_new} that RBF, Matern and Func4 produce very similar fits, whereas Poly2 tends to underfit the data.
This is also clear from the residuals (Fig.~\ref{fig:lrt_new}, lower panel), which are consistently larger for Poly2 than for the other three models. Interestingly, RBF, Matern and Func4 all show a slight spurious
bump where the models have been extrapolated across the signal window; outside of the signal window, deviations from zero are almost always within the error bars. Thus, visual inspection
rules out Poly2 but cannot be used to differentiate between RBF, Mattern, and Func4.


\begin{table}[htpb]
\centering
\caption{\textbf{Model comparison for background-only fits.} $|H|$ is the determinant of the Hessian of the marginal log-likelihood (Eq.~\eqref{GP:marginal}) evaluated at $\hat{\theta}$;
$n$ is the number of model parameters;
$d$ is the number of model degrees of freedom ($d = d_{\mathit{eff}}$ for GP regression, $d=n$ for Func4);
$\log(\text{PL})$ is the Poisson likelihood (Eq.~\eqref{pdg}) computed using either $f(x_i)$ evaluated at the optimal values of hyperparameters $\hat{\theta}$ from GP regression (Poly2,RBF,Matern) or found by non-linear maximization with $f(x_i)$ given by Eq.~\eqref{eq:func4} (Func4);
$\log(\text{GL})$ is the marginal log-likelihood (Eq.~\eqref{GP:marginal}). BIC (Eq.~\eqref{eq:bic:full}) and $\text{BIC}^\text{naive}$ (Eq.~\eqref{eq:bic})
were computed using $\log(\text{GL})$ for BIC and $\log(\text{GL})$/$\log(\text{PL})$ for $\text{BIC}^\text{naive}$.
AIC values (Eq.~\eqref{eq:aic}) were computed using $\log(\text{PL})$.
}  \label{kers}
\begin{tabular}{cccccccccc}
\hline \hline
Model     & $\log |H|$ & $n$ & $d$ &  -$\log(\text{PL})$  & -$\log(\text{GL})$  & $\text{BIC}^{\text{naive}}_\text{GL}$ & $\text{BIC}_\text{GL}$ & $\text{AIC}_\text{PL}$ & $\text{BIC}^{\text{naive}}_\text{PL}$ \\ \hline
Poly2      &  -0.531     & 1          & 2.99          & 38.02 & 87.52 &   89.22  & 87.25  & 82.02  & 39.72 \\
RBF        &   0.417     & 2          & 4.68          & 8.95   & 72.15 &   75.55  & 72.36  & 27.26  & 12.35 \\
Matern    &   2.906     & 2          & 5.67          & 8.69   & 72.30 &   75.70  & 73.75  & 28.72  & 12.09 \\ \hline
Func4     &   --            & 5          & 5               & 8.65   &  --        &   --        &  --        & 27.30  & 17.15 \\ \hline \hline 
\end{tabular}
\end{table}




\section{Signal extraction}

\begin{figure}[htpb]
    \centering
    \includegraphics[width=9.5cm]{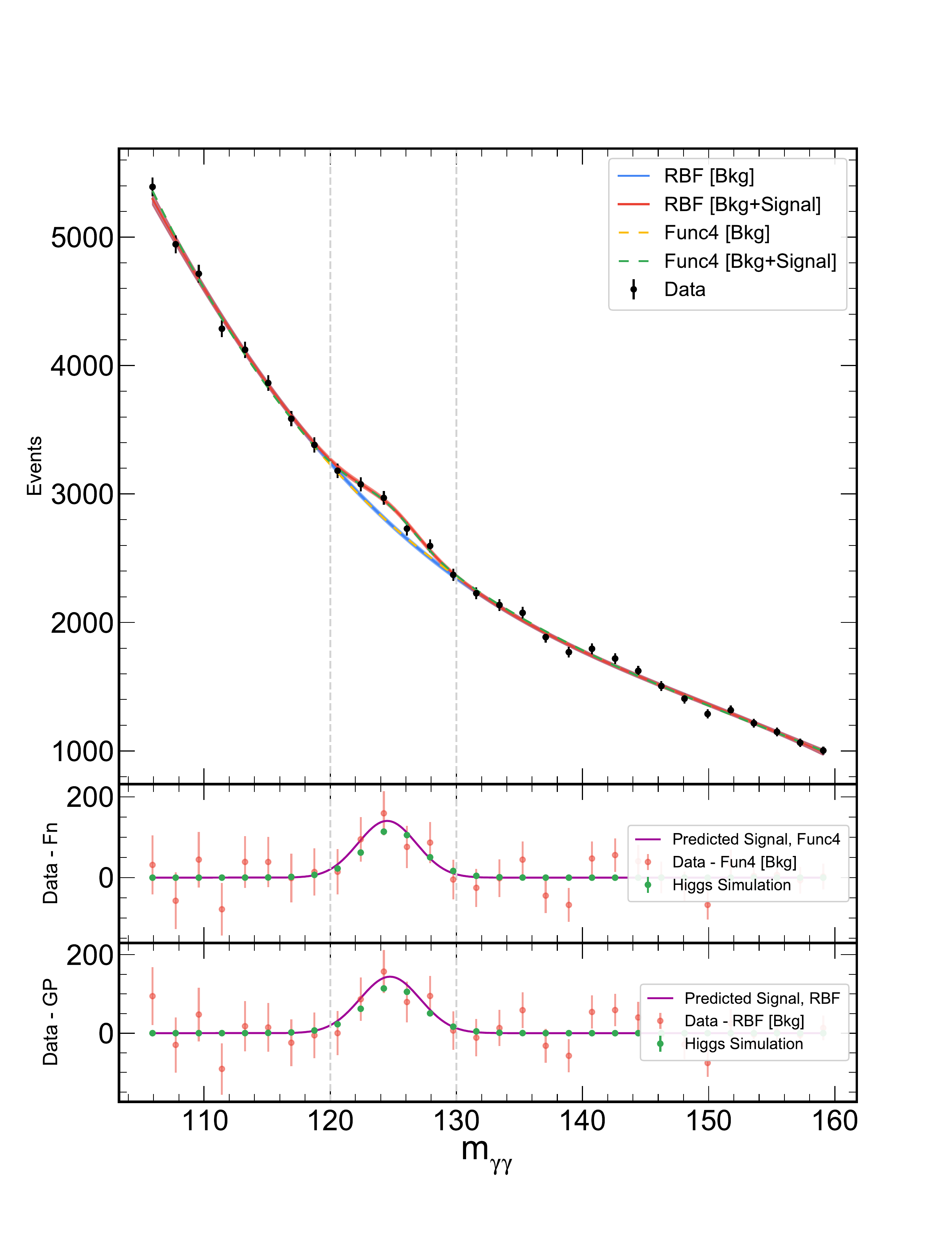}
    \caption{\textbf{Extracting the signal from the background distribution.} Upper panel: RBF[Bkg] and Func4[Bkg] are the background fits with the masked-out signal window (Fig.~\ref{fig:lrt_new}).
    RBF[Bkg+signal] and Func4[Bkg+signal] are the fits on the entire dataset, with kernel hyperparameters (RBF) and polynomial parameters (Func4) kept at their background-only values.
    Event counts $y_i$ are shown with error bars constructed using Garwood intervals (black markers). Middle panel: Gaussian signal predicted using Func4 background, residuals $y_i - f(x_i)$
    with respect to the Func4 background-only model, with $y_i$ error bars, and the results of the Higgs simulation described in Ref.~\cite{ATL-OREACH-PUB-2020-001}. Lower panel: same as the middle panel but with RBF instead of
    Func4. In all panels, vertical dashed lines indicate the boundaries of the signal window.
    }
    \label{higgs}
\end{figure}

In order to extract the signal superimposed on top of the background distribution, we have carried out GP regression with the RBF kernel and $m(x_i)$ given
by Eq.~\eqref{GP:meanf} using the entire dataset (Fig.~\ref{higgs}). 
Importantly, the kernel parameters were kept at the values $\hat{\theta}$ obtained via the previous fit to the background distribution, with the signal window masked out. Thus, the kernel hyperparameters
are responsible for modeling the background, while the Gaussian parameters in Eq.~\eqref{GP:meanf} are responsible for describing the signal. The resulting parameters are
$\{ A_\text{RBF}, \mu_\text{RBF}, \sigma_\text{RBF} \} = \{ 473 \pm 123, 124.7 \pm 0.6, 2.4 \pm 0.4 \}$. For comparison, we have also refit the
Func4 model (Eq.~\eqref{eq:func4}) on the entire dataset, by adding a Gaussian function (Eq.~\eqref{GP:meanf}) to
capture the signal contribution. As mentioned above, the Gaussian fitting parameters are
$\{ A_\text{Func4}, \mu_\text{Func4}, \sigma_\text{Func4} \} = \{ 443 \pm 199, 124.5 \pm 0.8, 2.3 \pm 0.9 \}$ in this case.
We observe that both RBF and Func4 are capable of capturing the approximately Gaussian signal which remains after subtracting the background distribution. On the basis of the
fitted Gaussian parameters, the correspondence between both models and between each model and the Higgs simulation (described in Ref.~\cite{ATL-OREACH-PUB-2020-001}) is overall very high (Fig.~\ref{higgs}).
However, we note that the GP approach with the RBF kernel extracts a clear Higgs signature consisting of $A_\text{RBF} = 473 \pm 123$ events above the background,
compared to $A_\text{Func4} = 443 \pm 199$ events from the Func4 fit. 
Thus, the mean number of predicted events is higher and the uncertainty is significantly lower with the GP RBF fit. The lower uncertainty of the prediction indicates that GP RBF is
preferable to Func4.




\subsection{Synthetic datasets for testing statistical significance of signal extraction}

To investigate the statistical significance of the observed signal, we have created 500 toy datasets based on the GP fit with the RBF kernel and $m(x_i) = 0$ to the background-only data.
This fit has generated an effective integer number of counts due to the background only, $N_\text{eff} = [ \sum_{i=1}^N f(x_i) ]$, where the square brackets indicate the rounding operation.
Next, we sampled from the GP predictive probability ${\cal N} (\tilde{y}_i|f(x_i),V(x_i))$, producing real-valued background ``counts'' $\tilde{y}_i$ in each bin $i$.
Finally, we used $\tilde{y}_i/\sum_{i=1}^N \tilde{y}_i$ as probabilities in a multinomial sampling process, generating a synthetic histogram of integer event counts. Each synthetic histogram was constrained to have
$N_\text{eff}$ counts, equal to the total number of events inferred to be due to the background. Note that our toy datasets include both the uncertainty inherent in GP regression and the uncertainty related to
generating integer event counts from the underlying model.
In order to create a full background+signal test set, we have added signal counts from the Higgs simulations~\cite{ATL-OREACH-PUB-2020-001} to each of the 500 background datasets.
Thus the signal component is fixed, while the background component varies from dataset to dataset according to the background model uncertainties.

\subsection{Test for biases in signal extraction}

To test the robustness of the fit, we check for potential biases in our background estimation procedure. Namely, for each of the 500 background+signal toy datasets described above,
we carry out a GP fit with the Gaussian mean function (Eq.~\eqref{GP:meanf}) on the entire dataset, while keeping the kernel hyperparameter values fixed at $\hat{\theta}$, values
found by the previously described fit to the background distribution, with the signal window masked out.
This procedure generates a set of predicted signal strength values, $\{ A_\text{pred} \}$, which can be compared with the corresponding exact value, $A_\text{true}$, the sum of
the event counts added to the background-only counts in order to create the combined background+signal toy datasets.
Specifically, we compute a Z-score like measure:

\begin{equation} \label{Zscore}
Z_W = \frac{A_\text{pred}-A_\text{true}}{\sigma_A},
\end{equation}
where $A_\text{pred}$ and $A_\text{true}$ are defined above and $\sigma_A$ is the standard deviation of the $\{ A_\text{pred} \}$ values.

\begin{figure}[htpb]
    \centering
    \includegraphics[width=7.5cm]{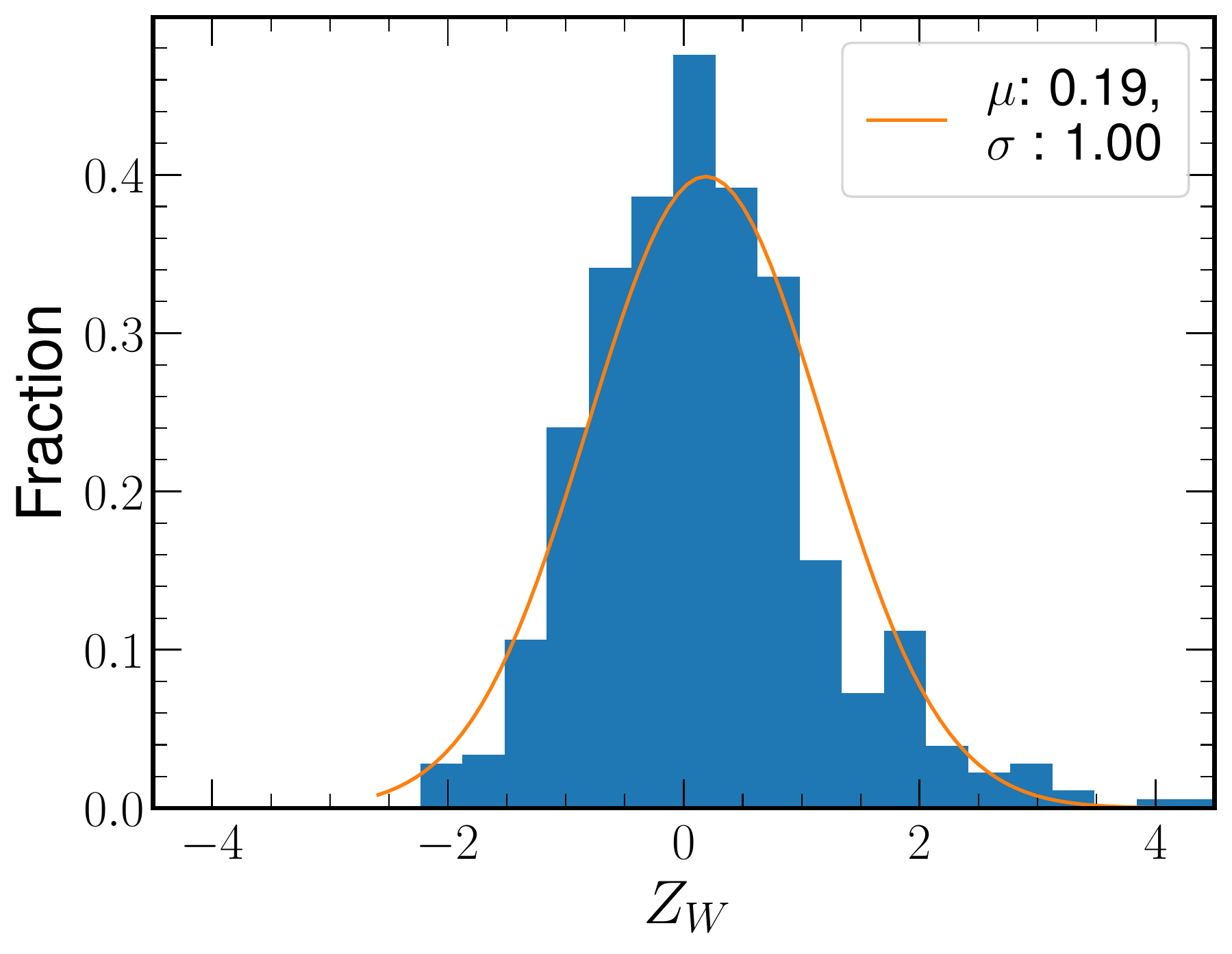}
    \caption{\textbf{Test for potential biases in signal extraction.} Shown is a normalized histogram of $Z_W$ scores (Eq.~\eqref{Zscore}) obtained by generating 500 toy datasets and carrying out GP regression as described in the text (blue bars). Orange curve: a Gaussian fit to the histogram, which yields $\mu=0.19$, $\sigma=1.00$. 
    }
    \label{bias}
\end{figure}

Fig.~\ref{bias} shows the resulting distribution of the $Z_W$ scores. We observe that the empirical distribution is well described by a Gaussian with $\mu=0.19$ and $\sigma=1.00$
(the latter is expected due to the normalization in Eq.~\eqref{Zscore}). The near-zero value of $\mu$ indicates that there are no substantial biases in our two-step background+signal reconstruction procedure.
We conclude that the signal contribution can be deconvoluted correctly from the underlying smooth background distribution.





\subsection{Posterior distributions of signal parameters and significance analysis}


We have investigated the posterior distributions of signal-characterizing parameters by carrying out Markov Chain Monte Carlo (MCMC) sampling~\cite{hasting} of the Poisson log-likelihood
(Eq.~\eqref{pdg}).
Routinely employed in Bayesian analysis, MCMC sampling of posterior probabilities is conceptually similar to studying model parameter sensitivity and estimating confidence
intervals in frequentist statistics~\cite{Feldman1998,Read_2002}.
Poisson rates $f(x_i)$ depend on the hyperparameter values $\hat{\theta}$ obtained via the previously described background-only fit and on the mean function $m(x_i)$, whose
parameters $\{ A, \mu, \sigma \}$ were sampled from the following priors: the prior for $A$ is uniform in the $[0, +\infty)$ range, while the prior for $\mu$ is Gaussian, with
the $124.7$ GeV mean and $0.02 \times 124.7$ GeV standard deviation. The $\sigma$ prior is also Gaussian, with the $2.4$ GeV mean and $0.1 \times 2.4$ GeV
standard deviation. The mean values are consistent with the Higgs simulation~\cite{ATL-OREACH-PUB-2020-001} and with the fits presented in Fig.~\ref{higgs}.
The $0.02$ and $0.1$ scaling factors in the priors are motivated by the ATLAS studies~\cite{Aad_2012}.
MCMC was implemented using the Emcee package~\cite{emcee} (\texttt{https://emcee.readthedocs.io}), with $10^4$ samples in each of 12 independent MC trajectories.

\begin{figure}[ht]
    \begin{subfigure}{.5\textwidth}
    \centering
    \includegraphics[width=0.8\linewidth]{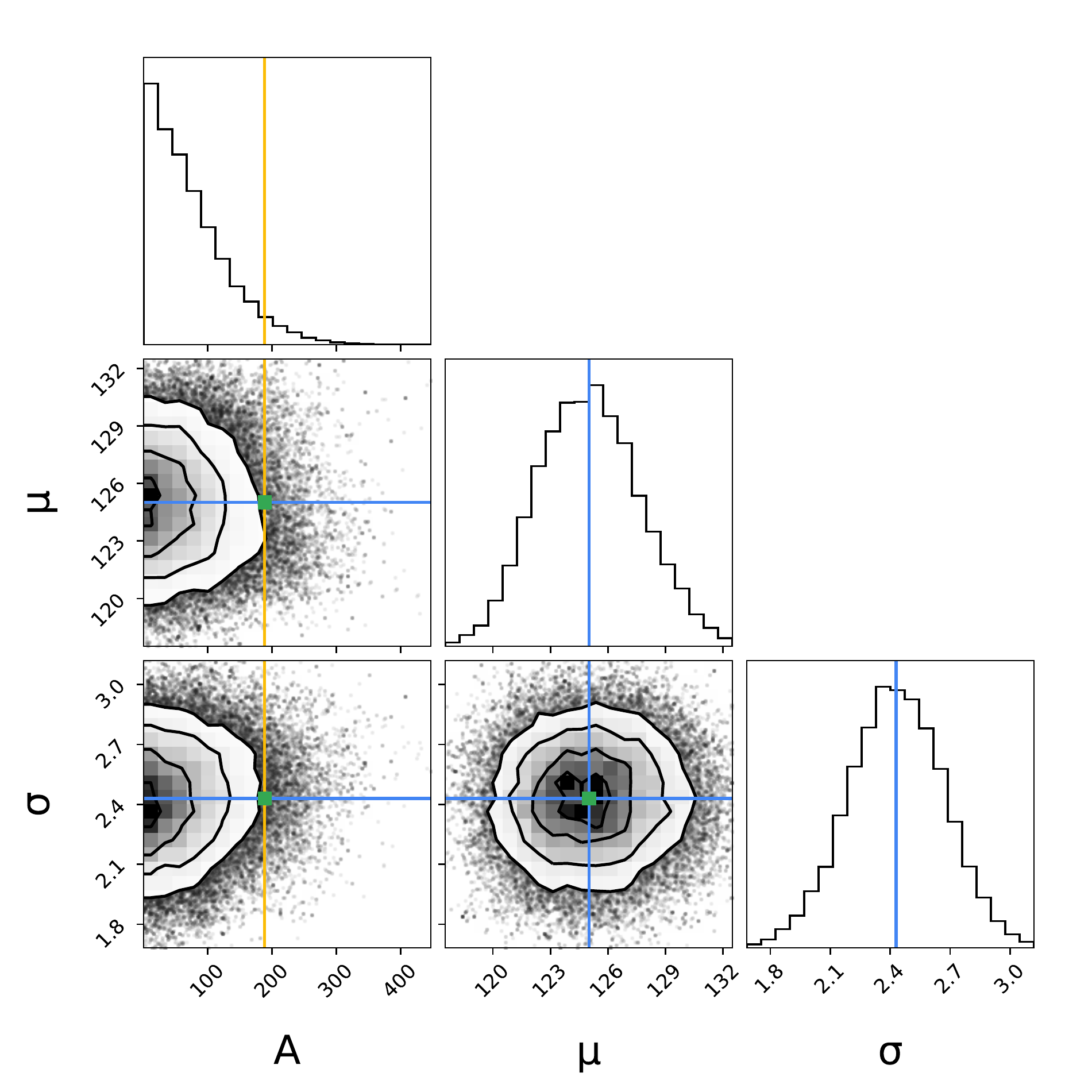}
    \caption{}
    \end{subfigure}
    \begin{subfigure}{.5\textwidth}
     \centering
    \includegraphics[width=0.8\linewidth]{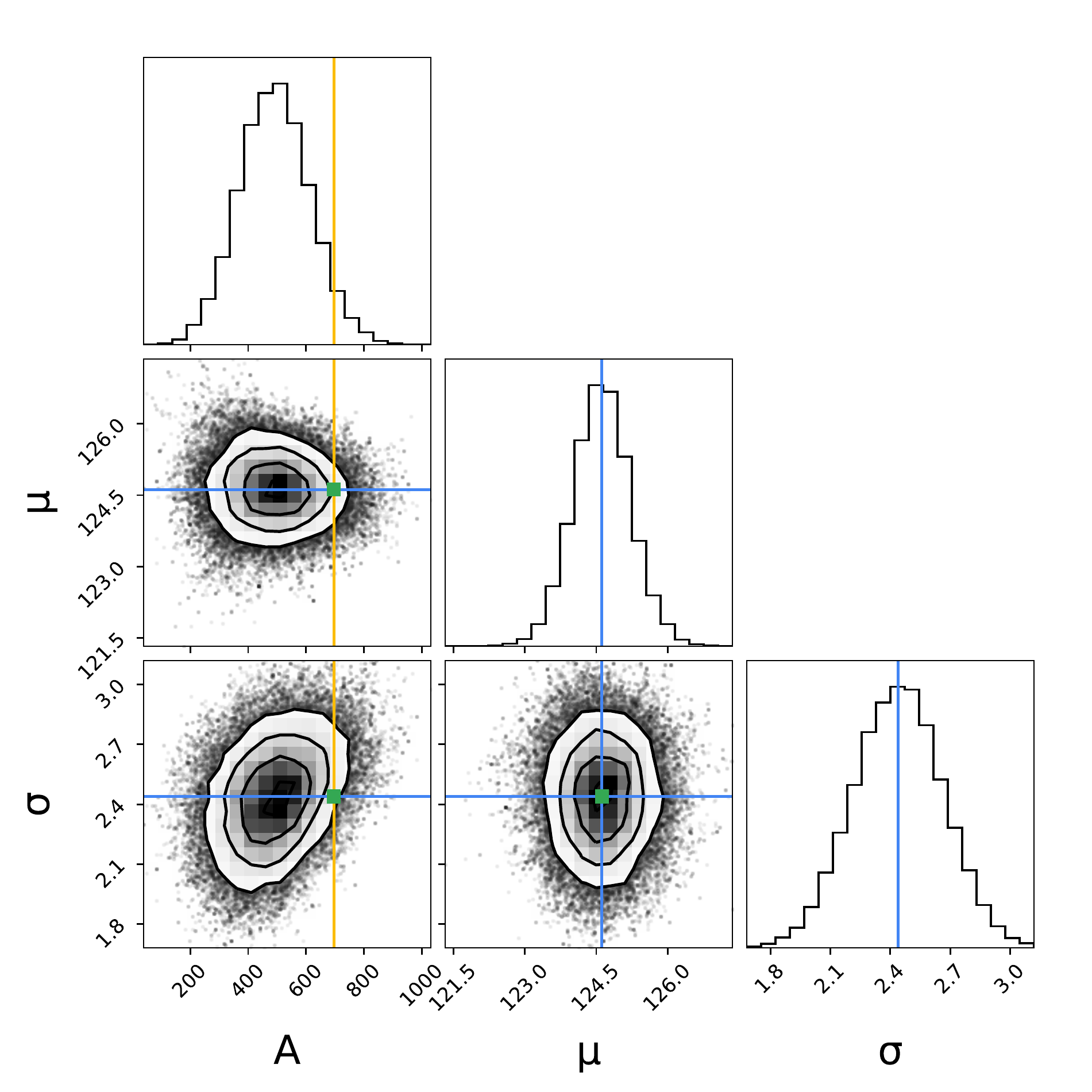}
    \caption{}
    \end{subfigure}
    \caption{\textbf{Posterior distributions of signal parameters obtained by MCMC sampling.} Shown are posterior probabilities in the corner plot format
    (\texttt{https://corner.readthedocs.io/en/latest/})~\cite{corner}. Diagonal entries: marginalized posterior probabilities $P(A)$, $P(\mu)$, $P(\sigma)$, off-diagonal entries: joint posterior probabilities
    $P(\mu,A)$, $P(\sigma,A)$, $P(\mu,\sigma)$, with the contours indicating 2D sigma levels. 
    Blue lines mark the median values in the posterior distributions, yellow lines mark
    the 95\% quantile values. Panel (a): MCMC analysis of a randomly chosen synthetic dataset (out of 500 toy datasets with background counts only). 
    Panel (b): MCMC analysis of the observed counts, which include both background and signal.
    }
    \label{posterior}
\end{figure}

Fig.~\ref{posterior} shows MCMC posterior distributions of the three parameters characterizing the signal: overall signal strength $A$, the mean position of the signal peak $\mu$, and the width of the signal
peak $\sigma$. In Fig.~\ref{posterior}a MCMC sampling was based on a synthetic dataset without any signal added, which was randomly chosen among the 500 background-only test sets described above.
As expected, $P(A)$, the marginalized posterior probability for signal strength, is highest when $A$ is close to zero and falls off rapidly as $A$ increases, while $P(\mu)$ and $P(\sigma)$
appear Gaussian. Moreover, the correlations between all 3 parameter pairs appear to be weak. In contrast, when the real data is analyzed which contains both the background counts and Higgs events,
the maximum posterior probability value of $A$ is located around 500 counts, consistent with the earlier Hessian analysis of GP regression with the RBF kernel (Fig.~\ref{posterior}b).
Indeed, a Gaussian fit of $P(A)$ in Fig.~\ref{posterior}b has yielded $485 \pm 121$ Higgs events, very close to the $473 \pm 123$ Higgs events obtained earlier using the GP regression framework.
Thus there is a clear signature of Higgs counts in the real data. Interestingly, the joint probability $P(\sigma,A)$ reveals a correlation between signal strength and signal width, with stronger signals tending to have larger widths.

\begin{figure}[ht]
    \centering
    \includegraphics[width=7.5cm]{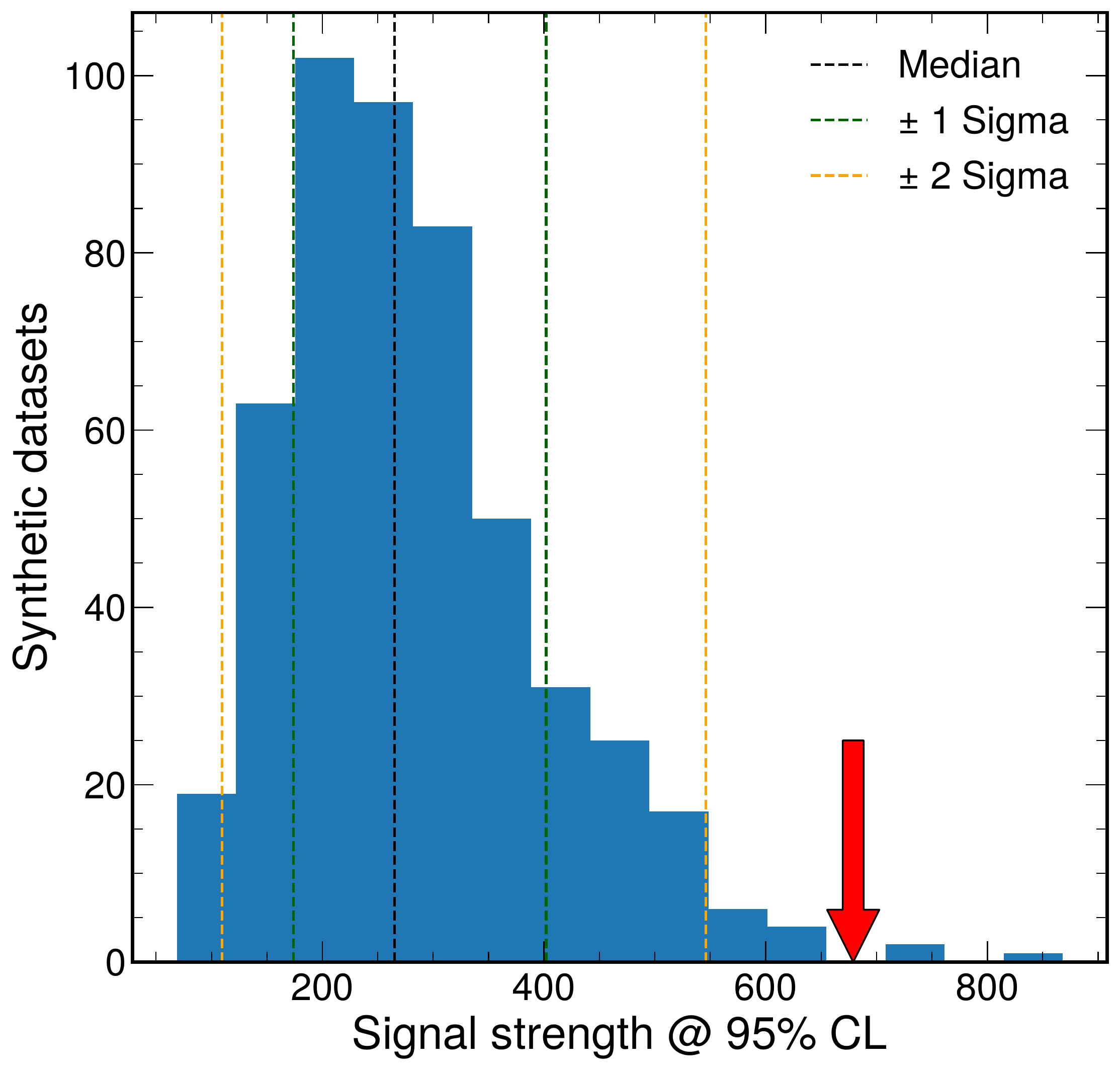}
    \caption{\textbf{Distribution of the signal strength $A$ in background-only datasets.} Shown is the histogram of the 95\% quantiles (confidence levels, or CL) for $A$ inferred from 500 synthetic datasets
    with background-only counts using MCMC sampling. Black dashed line indicates the 50\% quantile, or the median, of the 95\% CL distribution
    (i.e., the median number of Higgs events observed above background, reported at 95\% CL). From left to right, the yellow lines show 2.5\% and 97.5\% quantiles and the
    green lines show 16\% and 84\% quantiles, respectively. The red arrow indicates the 95\% CL value obtained from the data (cf. Fig.~\ref{posterior}b).
    }
    \label{limit}
\end{figure}

To provide a more quantitative estimate of the statistical significance of the signal strength observed in real data, we have plotted a histogram of the 95\% confidence levels for $A$ for all 500
background-only toy datasets (Fig.~\ref{limit}). The value observed with the actual data is 3.02 $\tilde{\sigma}$ above the median, where $\tilde{\sigma}$ is the distance between
the median and the 84\% quantile, and is larger than 99.4\% of the values empirically observed in the histogram.

\section{Summary}

In this work, we have developed a procedure for using Gaussian Process (GP) regression to extract localized signals from smooth background distributions. Although this procedure is of interest in many
areas of science, including astrophysics and crystallography, here we focus on extracting Higgs events from an ATLAS open dataset which consists of binned event counts. Despite its relatively
small size, this is a challenging dataset since the putative signal is masked by the background and the inference procedure used to analyze the data affects the statistical significance of the predicted signal.
Traditionally, the background distribution is modeled using a polynomial fit, onto which a Gaussian signal is superimposed (Eq.~\eqref{eq:func4})~\cite{ATL-OREACH-PUB-2020-001}.
Here we propose an alternative framework in which GP regression is used to model the background and the signal is modeled via the mean function, which affects both the marginal likelihood
of the observed data (Eq.~\eqref{GP:marginal}) and the predictive probability, the conditional probability of a new datapoint given the previously observed data (Eq.~\eqref{mean:var:general}).
As with the functional fits, the mean function is represented by a Gaussian with three free parameters (Eq.~\eqref{GP:meanf}), one of which, $A$, is especially relevant to us since it represents 
the total number of signal events found in the dataset.

The GP framework is more flexible than more standard approaches which employ a fixed set of basis functions such as polynomials or Gaussians~\cite{bishop2006pattern,gpml}. This flexibility comes from the focus of the GP-based approach on the correlation structures in the dataset, which are modeled using kernel functions.
Although GP methods are not limited by the prior choice of the finite basis (indeed, some popular kernels correspond to infinite basis sets) and the GP approach is in principle fully
Bayesian, most kernel functions depend on one or several hyperparameters, such as the characteristic length scale in the RBF kernel. A fully Bayesian treatment of the dependence
of the model evidence on hyperparameters is usually impractical; however, simply maximizing the evidence with respect to hyperparameters may lead to overfitting for more complex kernels.
In order to provide a more principled approach to the selection of the kernel type, we have considered two independent methodologies.

One of them, BIC, is based on evaluating the model
evidence under the Laplace approximation and the assumption that the effects of hyperparameter priors are negligible (Eq.~\eqref{eq:bic:full}). With several additional approximations, notably
the assumption that the Hessian matrix has full rank, BIC yields a simple correction which penalizes model complexity (Eq.~\eqref{eq:bic}). The other approach, AIC,
is non-Bayesian. Instead of concentrating on the model evidence, it focuses on the degree of smoothing that results from applying a given kernel to the dataset. Thus, the AIC and BIC approaches
are complementary and reflect different kernel properties (amount of data smoothing vs. the shape of the log-likelihood landscape as a function of hyperparameters). Using both criteria
holistically, we have chosen a well-known RBF kernel for our GP regression models, although the results with the Matern kernel are only slightly worse.

We note that AIC yields approximately equal scores for the GP RBF fit and the traditional fit, which models the background using a fourth-order polynomial (Table~\ref{kers}).
The results of the two fits are visually very similar when the GP mean predictive probability is compared with the ML curve produced by the functional fit, and both
approaches are close to the Higgs simulations predictions (Figs.~\ref{fig:lrt_new},\ref{higgs}).
However, the total area $A$ under the signal bump is somewhat higher with the GP RBF fit compared to the functional fit, with $473$ and $443$ Higgs events, respectively. 
More critically, Hessian analysis reveals that the standard deviation is much smaller with the GP prediction, $123$ vs. $199$ in the functional fit. Thus, the GP approach is
preferable since it leads to the higher signal strength prediction with considerably less uncertainty.

After ascertaining that our signal extraction procedure is not biased (Fig.~\ref{bias}), we have proceeded to investigate the posterior probabilities of model parameters by MCMC sampling (Fig.~\ref{posterior}). This computational approach is necessary since we have focused on the Poisson log-likelihood (Eq.~\eqref{pdg}), which is more appropriate for modeling integer event counts. The Poisson log-likelihood depends on the kernel hyperparameters, which were kept fixed to their values $\hat{\theta}$ obtained by fitting to the background-only data (Fig.~\ref{fig:lrt_new}),
and on the signal strength, mean and width, which were sampled from prior distributions. The prior for signal strength $A$ was uninformative, assigning equal weights to any non-negative value.
The priors for the mean and the width were informative, modeled by Gaussians whose parameters were constrained by Higgs simulations (Fig.~\ref{higgs}) and by the studies of instrumental errors in the
ATLAS detector~\cite{Aad_2012}. The resulting posterior probability for signal strength shows a clear Higgs signature, with $485 \pm 121$ Higgs events (Fig.~\ref{posterior}b).
These numbers are consistent with the previous estimate obtained by Hessian analysis of the signal parameters in GP regression, which yielded $473\pm123$ Higgs events.

When the MCMC sampling procedure is applied to synthetic datasets where no contributions from the signal are expected, the posterior distribution for signal strength is centered on zero and
the typical predicted values are much smaller (Fig.~\ref{posterior}a). The latter is clearly seen by combining the data from $500$ independently generated background-only synthetic datasets into a histogram
of 95\% confidence levels for signal strength $A$ (Fig.~\ref{limit}). The corresponding confidence level obtained from the real dataset is larger than $99.4\%$ of the histogram values and
corresponds to $3.02~\tilde{\sigma}$, where $\tilde{\sigma}$ is the distance between the median and the 84\% quantile. Thus our signal strength prediction
is also highly significant within the MCMC framework.

In summary, we have developed a novel GP regression framework for extracting localized signals from smooth background distributions of unknown functional form. This problem appears in many
areas of science where a weak signal of interest is masked by background events due to light scattering, extraneous emission sources, etc. The location and the width of the signal can
sometimes be guessed based on physical considerations; in other cases, consideration of multiple putative signal windows is necessary,
as in LHC anomaly detection searches \cite{ad1,ad2,ad3,ad4,ad5,ad6}. 
In both scenarios, only rough estimates of the position and the width of the signal window are required.
Data outside of the signal window is assumed to belong to the background and a GP model without the signal
contribution is fitted to it. We carry out model selection using both BIC and AIC considerations, including an in-depth analysis of the BIC assumptions.
The extrapolation of the model across the signal window
then provides an estimate of the background, from which the signal can now be separated in a second GP fit where only
the signal parameters are allowed to vary, while all the background parameters remain fixed. This two-step procedure allows us to deconvolute the signal from the background in a robust
and reproducible manner. An application of our approach to the open Higgs boson dataset from the ATLAS detector (known as the ATLAS open dataset)
yields a highly significant prediction of the Higgs boson signature, outperforming the traditional approach based on fitting a polynomial function to the background distribution.

\clearpage

\acknowledgments

This work was supported by the National Science Foundation through grants NSF-PHY-1607096 (to A.L.)
and NSF-MCB-1920914 (to A.V.M.). This manuscript has been authored by Fermi Research Alliance, LLC under Contract No. DE-AC02-07CH11359 with the U.S. Department of Energy, Office of Science, Office of High Energy Physics.


\bibliographystyle{jhep}
\bibliography{main}
\end{document}